\documentclass[aps,prl,reprint,superscriptaddress,noeprint]{revtex4-2} 

\usepackage{amsmath}
\usepackage{amssymb}
\usepackage{wasysym}
\usepackage{graphicx}
\usepackage{hyperref}
\usepackage{xcolor}
\usepackage{bm}  
\usepackage{orcidlink}
\usepackage{physics2}
\usephysicsmodule{ab, ab.braket}  
\usepackage{siunitx}  
\DeclareSIUnit\gauss{G}
\DeclareSIUnit\bohr{\ensuremath{a_0}}
\usepackage{csquotes}
\usepackage{braket}
\usepackage{placeins} 

\begin{document}

\title{Isothermal compression of a Fermi gas to deep quantum degeneracy}

\author{Kirill Karpov${}^{\orcidlink{0009-0000-2701-5593}}$}
\author{Jonas Auch${}^{\orcidlink{0009-0005-5863-8239}}$}
\author{Eduard Heidt${}^{\orcidlink{0009-0007-1294-5043}}$}
\author{Florian Kiesel${}^{\orcidlink{0009-0009-0515-3069}}$}
\author{Alexandre De Martino${}^{\orcidlink{0009-0008-6358-7503}}$}
\affiliation{Physikalisches Institut, Eberhard Karls Universit\"at T\"ubingen, 72076 T\"ubingen, Germany}
\author{Christian~Gro\ss${}^{\orcidlink{0000-0003-2292-5234}}$}
\email{christian.gross@uni-tuebingen.de}
\affiliation{Physikalisches Institut and Center for Integrated Quantum Science and Technology, Eberhard Karls Universit\"at T\"ubingen, 72076 T\"ubingen, Germany}

\date{\today}

\begin{abstract}
	The standard approach for generating deeply degenerate quantum gases is evaporative or sympathetic cooling in a harmonic trap, after which the gas has reached its minimum entropy.
	All subsequent state transformations rely on adiabatic changes of a closed system, and coupling to the environment or non-adiabatic processes monotonically increase the entropy.
	Here, we demonstrate that this experimental paradigm can be bypassed by utilizing species-selective trapping with a low-dissipation optical tune-out trap in a dual-species mixture.
	We successfully reduce the entropy of a two-component fermionic quantum gas via isothermal compression within a bosonic bath, reaching deep quantum degeneracy of $T/T_F = \num{0.024}^{+0.007}$, with $T_F$ the Fermi temperature.
	By characterizing the cross-dimensional relaxation and thermalization, we demonstrate that cooling light fermions with heavy bosons remains efficient and fast, even deep in the degenerate regime, where the thermalization time is found to be independent of $T/T_F$.
	Our results pave the way for direct cooling within optical lattices, box traps, or other complex potentials, thereby eliminating the reliance on adiabatic state transformations to reach strongly interacting many-body regimes.
\end{abstract}

\maketitle

Ultracold Fermi gases provide a powerful platform for exploring and simulating strongly correlated many-body physics~\cite{bloch_08}.
Experimental advances in cooling have unlocked a diverse range of many-body phenomena, among them the realization of the BEC-BCS crossover in the continuum~\cite{zwerger_12} and lattice phenomena at lower and lower entropies in the Hubbard model~\cite{gross2017-cg}, from the Mott insulator~\cite{jordens_08,schneider_08} via antiferromagnetism ~\cite{greif_13,hart_15,hilker2017-cga,shao_24,xu_25} to collective physics in the doped regime~\cite{hartke2025-cg,chalopin2026-cg, kendrick2025-cg}.
Despite this rapid experimental progress, the minimally achievable entropy remains the primary limiting factor for observing novel phases, such as the elusive Fulde-Ferrell-Larkin-Ovchinnikov (FFLO) state~\cite{fulde_1964,larkin_1964,radzihovsky_2010} and $d$-wave superconductivity~\cite{bohrdt2021-cg,hofstetter_02,keimer_15}.

The restriction to standard evaporative and adiabatic techniques inherently limits the lowest possible entropy.
While the phase-space density remains strictly constant in a perfectly adiabatic process, this ideal situation cannot be realized experimentally.
Imperfect system isolation limits the available timescales, such that the experimental optimum entropy is always a trade-off between heating and non-adiabatic transformations.
In contrast, in an isothermal process, the absolute temperature remains constant, anchored by the bath, and low entropies can be reached if a cold enough bath is available.

Implementing such isothermal transformations in ultracold atomic systems requires both efficient coupling to the thermal bath and independent control over system and bath energy scales.
We achieve both requirements in a mixture of bosonic $^{166}$Er and fermionic $^6$Li: efficient coupling by relying on the large scattering cross section, and energy scale control by differential trapping.
A major experimental advantage of heteronuclear mixtures lies in their distinct optical properties, which enable the realization of species-selective potentials.
This is optimally achieved by employing a tune-out wavelength, where the dynamic polarizability of one species vanishes~\cite{leblanc_07}.
However, implementing these traps typically introduces severe off-resonant photon scattering for at least one of the components~\cite{catani_09,vaidya_15,lippi_24,martino_25}.
The resulting heating has so far been the limiting factor for enhanced cooling.
In contrast, for our mixture, we make use of a low-heating tune-out wavelength of Er at \qty{841}{\nano\meter} for independent optical trapping of Li~\cite{martino_25}.

Here we demonstrate an isothermal cooling protocol for fermions in a bosonic bath, leading to one of the deepest degenerate fermionic systems realized so far~\cite{hadzibabic_03, delehaye_15}, with $T/T_F = \num{0.024}^{+0.007}$.
We show that the fermion-to-bath collisions, crucial for the isothermal process, remain effective down to the lowest reduced temperatures of a few percent of the Fermi temperature $T_F$: the interspecies thermalization time is independent of the degeneracy, since Pauli blocking of the collisions is compensated by the reduced fermionic heat capacity.
These results are enabled by a novel heavy-light Bose-Fermi mixture~\cite{baroni_24}, for which we achieve simultaneous quantum degeneracy, demonstrate the collisional stability, and measure the background scattering length.
Our results, obtained for a spin mixture of non-interacting fermions, extend straightforwardly to any spin imbalance and any trapping configuration, and leave room for improvement via the many Feshbach resonances of the ErLi mixture~\cite{schafer2022-cg} to optimize the collisional properties for faster cooling.

\begin{figure}
	\includegraphics{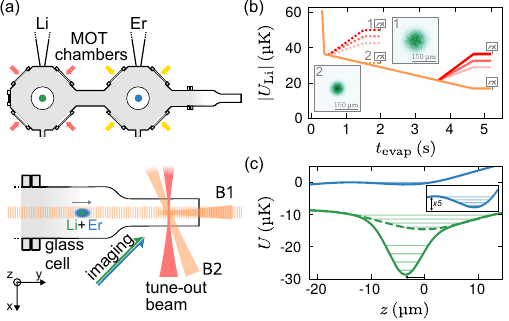}
	\caption{Experimental setup and compression of Li.
		(a) Vacuum system and ODT setup.
		The vacuum system consists of spatially separated MOT chambers for Er (blue) and Li (green), each with its own Zeeman slower.
		After the first cooling steps, Er and Li are transported together to a high optical access glass cell by a running lattice.
		B1 propagates in the $y$ direction and is crossed by the almost perpendicularly oriented B2.
		Additionally, the tune-out beam and the imaging beams are indicated.
		(b) Dual-species evaporation trajectory and $T_F$ tuning by compression with the tune-out beam (red).
		The $\qty{1064}{\nano \meter}$-trap depth is decreased by two linear power ramps (orange).
		To perform the isothermal compression, we ramp the tune-out beam power to different values over \qty{1}{\second} during the evaporative cooling stage, followed by a \qty{0.5}{\second} hold time to ensure full thermalization.
		To probe different temperatures, we stop the evaporation ramp at different points and, optionally, start the tune-out beam ramp \qty{1.5}{\second} before the measurement (dashed lines, higher temperature setting).
		ToF absorption images of Li without (1) and with (2) isothermal compression reveal the changed Fermi temperature/velocity directly through the further expanded cloud.
		(c) Trap depth of Er and Li at the final evaporation point.
		For Er (blue) the trap depth is heavily influenced by gravity, resulting in a tilted potential and a shifted trap center.
		Due to the small influence of gravity on Li, the potential for Li in the same trap configuration of B1 and B2 (green, dashed) is about 20 times deeper.
		The tune-out beam provides independent control of the center of the Li trap and further increases the trap depth for Li (green, solid).
		The energies of the 10$^\text{th}$, 30$^\text{th}$, 50$^\text{th}$, and 70$^\text{th}$ harmonic-oscillator eigenstates, shown for all traps, visualize the increased energy scale in the tune-out beam.
	}
	\label{fig:fig1}
\end{figure}

Our experiment begins by simultaneously loading spatially separated magneto-optical traps (MOTs) of bosonic $^{166}$Er and fermionic $^6$Li, from which both species are transported together by a running optical lattice into a glass cell, where all reported measurements are performed (see \autoref{fig:fig1})~\cite{supps}.
There, the mixture is reloaded into two horizontally oriented crossed single-mode laser beams of \qty{1064}{\nano\meter} light, beam one (B1) and beam two (B2), followed by an intermediate evaporation stage.
For the final evaporation, a vertical magnetic offset field of about \qty{1.3}{\gauss} is set to minimize three-body losses for Er~\cite{krstajic_23}.
We work with Er in the energetically lowest state $\ket{J=6,\,m_J=-6}$ and with Li in a balanced spin-mixture of $\ket{F,m_F}=\ket{1/2,1/2}\equiv\ket{1}$ and $\ket{1/2,-1/2}\equiv\ket{2}$.
At this magnetic field, the Li $\ket{1}, \ket{2}$ spin-mixture is non-interacting~\cite{zuern2013}.

The evaporation is performed by two linearly decreasing power ramps of B1 and B2, reducing the trap depth for Li, $U_\text{Li}$, from \qty{178}{\micro\kelvin} to \qty{15}{\micro\kelvin} within \qty{4.7}{\second}~\cite{supps}.
Since the polarizability for \qty{1064}{\nano\meter} light of Er is a factor of $0.6$ lower than that of Li~\cite{becher_18, safronova}, only Er is evaporated out of the trap while the Li atom number stays constant~\cite{supps}.
At the end of the evaporation, we achieve a double-degenerate system of Er and Li with a detection-limited temperature of $T_\text{Li}/T_F=0.14$ ($0.06$ when using the Er temperature as a thermometer) of the \num{11e3} Li atoms and observe a Bose-Einstein condensate (BEC) of \num{1.3e3} Er atoms, with a condensate fraction of \qty{82}{\percent}~\cite{supps}.
Further evaporation in this single-wavelength optical dipole trap (ODT) is limited by overlap loss from gravitational sagging of (heavy) Er and by the opening of the trapping potential for Er to its final depth $U_{\text{Er}}$ by gravity.

\begin{figure}
	\includegraphics[scale=1]{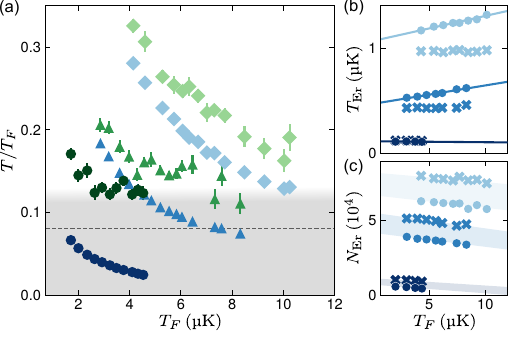}
	\caption{Isothermal compression at three different initial temperatures.
	(a)~Temperature of the fermions measured directly (green) and via the bosons (blue).
	The three initial temperatures ($T_\text{Er} = {}$\qtylist{1.1;0.5;0.1}{\micro\kelvin}) are marked by the decreasing brightness of the color and additionally by the symbol.
	For the bosonic measurements, we normalize to the $T_F$ directly measured by the ToF expansion of the fermions.
	The fade-in of the gray shading indicates the increasing systematic effect on the direct temperature extraction, with the dashed line marking the temperature we would extract for a zero-temperature gas.
	(b) Er bath temperature $T_{\text{Er}}$ measured with (circles) and without (crosses) the Li gas.
	The temperature increase in the Er bath, induced by the compression of Li, is visibly suppressed at lower initial temperatures.
	Solid lines denote linear fits to the compression data~\cite{supps}.
	(c) Er bath atom number $N_{\text{Er}}$ measured with (circles) and without (crosses) Li. In all regimes, trap compression induces a slight downward trend in the Er atom number with a minor dependence on the presence of Li.
	The most pronounced effect of the presence of Li is a constant offset in the atom number from its extra heat load early in the evaporation.
	Error bars denote the standard error of the mean.
	}
	\label{fig:fig2}
\end{figure}

To reach deeper degeneracies, we implement a species-selective trap for Li using a frequency-stabilized Ti:Sa laser tuned to an Er tune-out wavelength at approximately \qty{841}{\nano\meter}, detuned by $\qty{244}{\giga\hertz} \approx \num{3.1e7}\,\Gamma_{841}$ from the nearby narrow resonance of linewidth $\Gamma_{841}=\qty{8}{\kilo\hertz}$~\cite{ban_05, martino_25, supps}.
While ideally not affecting Er, this elliptical beam provides a far-detuned attractive ODT for Li, dominating the confinement in the $y$- and $z$-directions.
Importantly, the position of the Li cloud can be controlled freely, which allows us to perfectly compensate the gravitational sag of Er, and thus maintain the overlap of the clouds.
To load the Li atoms into the tune-out beam, we increase the \qty{841}{\nano\meter} laser power within \qty{1}{\second} during the evaporation, followed by \qty{0.5}{\second} of holding.
As we show below, these times are longer than the interspecies thermalization time, ensuring that the mixture remains in thermal equilibrium: $T_\text{Er}\approx T_\text{Li}$.
The trap depth for Li, $U_\text{Li}$, is now dominated by the tune-out beam, and we use this independent control to tune the Fermi temperature.
In the absence of Er, ramping up the tune-out beam provides an isentropic compression of Li, leaving the entropy, measured by $T/T_F$, unaffected.
However, when Er is present, the compression of Li becomes isothermal and increases the degeneracy by decreasing $T/T_F$.
In \autoref{fig:fig2} we show the evolution of $T/T_F$ and $T_\text{Er}$ as $T_{F}$ is increased by the tune-out beam, for three different initial temperatures.
For Li we find $T/T_F$ from the 2D polylogarithmic fit~\cite{ketterle_08}.
For Er we extract $T_\text{Er}$ from time-of-flight (ToF) expansion measurements for temperatures larger than the critical temperature and from the condensate fraction in the degenerate regime~\cite{ketterle_99}.
ToF thermometry for fermions critically depends on the resolution of the sharp features in the wings of the low-temperature cloud.
Various noise sources broaden these in our experiment, affecting the directly measurable temperature more strongly for lower $T/T_F$; we estimate that a $T=0$ Fermi gas would be measured as $T/T_F \approx 0.08$~\cite{supps}.
Thus, we use sympathetic thermometry by the bosonic component for low $T/T_F$~\cite{lous_17}.
As a reliable test of the thermal contact, we exploit a quench-induced cloud anisotropy of Li, which relaxes only via collisions with the bath and persists in its absence, as demonstrated below.
For the coldest starting point of the compression, bosonic thermometry reveals a reduced temperature at the strongest compression of $T/T_F = \num{0.024}^{+0.007}$, where the upper bound is an estimate for the Li temperature taking heating by light scattering at maximum tune-out beam power and the measured thermalization time into account~\cite{supps}.
In the harmonic trap, this corresponds to an entropy per particle of $S/N = \pi^2 k_B\, T/T_F \approx 0.24\,k_B$.
Neither of these temperature values is a fundamental limit.
In the present configuration, the Er trap depth at the end of the evaporation is limited by gravity (see \autoref{fig:fig1}(c)).
Further lowering the bath energy scale, and with it the bath temperature, would require, for example, magnetic levitation of Er or transfer to a trap with larger aspect ratio.

Next, we explore the reaction of the bath to the compression of the fermions (see~\autoref{fig:fig2}(b,c)).
The compression only deepens the Li trap, so no Li atoms are lost.
In the absence of Li, the Er temperature remains constant when the tune-out beam is ramped up, confirming a negligible direct heat load.
We observe a slight, heating-free atom loss which we attribute to the photon scattering of Er and discuss further in the Supplemental Material~\cite{supps}.
With Li, the Er temperature increases due to the energy flow from the compressed fermions, while the atom-number dynamics are unchanged up to a constant offset (see caption of \autoref{fig:fig2}).
The temperature increase is moderate and becomes non-measurable at the coldest initial temperatures:
it is on the order of \qty{15}{\percent} for an initial $T/T_F \approx 0.3$, while at the same time the fermion degeneracy increases by a factor of three.
This lower reaction at higher degeneracy reflects the decreasing heat capacity of the fermionic gas, which is carried by the thermally active fraction $\propto T/T_F$ of the fermions around the Fermi energy $E_F$~\cite{presilla_03}.
Thus, at high initial temperatures the compression is near-isothermal, while at low ones, it becomes fully isothermal.
Overall, the mixture shows all the expected thermodynamic signatures characteristic of the isothermal compression process.

\begin{figure}
	\centering
	\includegraphics{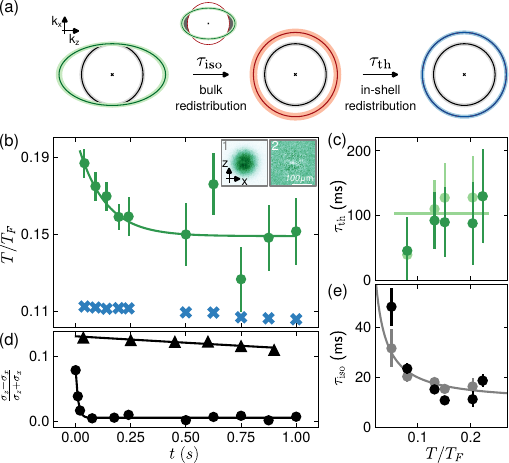}
	\caption{
		(a) Sketch of the isotropization and thermalization processes in a 2D situation (for better illustration), to scale for $T/T_F \approx 0.1$.
		The initial Fermi sea in the $k_x, k_z$ plane is illustrated by the chemical potential $\mu$ (black circle) surrounded by the thermally active shell (lighter region); the cross marks the origin.
		The quench induces an anisotropic chemical potential (green), lifting a large fraction of the Fermi sea (dark gray in the smaller upper picture) above the isothermal equilibrium chemical potential (red); this fraction is redistributed between the directions within $\tau_\text{iso}$, with Pauli blocking effective in it.
		The subsequent thermalization to the bath's temperature (the thermally active shell of the final state (blue) has the same width as the initial one) proceeds largely within the thermally active shell; thus, $\tau_\text{th}$ is independent of $T/T_F$.
		(b) Temperature relaxation of Er (blue) and Li (green) following a quench on the Li gas.
		The green solid line is an exponential fit to the data used to extract $\tau_{\text{th}}$.
		The inset shows the Li density distribution after ToF (1) and the residual to a polylogarithmic fit (2) to highlight the Er-induced density depression developing during ToF~\cite{supps}.
		(c) Interspecies thermalization time $\tau_{\text{th}}$ (dark green) of the ErLi mixture as a function of the equilibrium degeneracy parameter $T/T_F$.
		Light green points denote $\tau_{\text{th}} \frac{\bar{n}}{\bar{n}_0}$, representing the thermalization time corrected for variations in the interspecies density overlap.
		The horizontal line indicates the mean value of $\tau_{\text{th}}$.
		(d) Relaxation of the spatial anisotropy of the Li cloud with (points) and without (triangles) the Er bath.
		Solid lines are exponential fits to the data used to extract $\tau_{\text{iso}}$.
		(e) Isotropization time $\tau_{\text{iso}}$ (black) of Li as a function of $T/T_F$.
		Gray points display the isotropization time corrected for changes in the density overlap.
		The gray line shows a $(T/T_F)^{-1}$ scaling, illustrating the increase in isotropization time at low temperatures.
		In (b) and (d) the error bars denote the standard error of the mean; in (c) and (e) the standard uncertainties of the fit parameters.
	}
	\label{fig:fig3}
\end{figure}

Maintaining thermal equilibrium at all temperatures is most critical for our technique, and this has been discussed as a major challenge for deep fermion cooling~\cite{presilla_03, onofrio_16}, even more so for a strongly mass-imbalanced mixture such as ours~\cite{mudrich_02, silber_05, marzok_07, desalvo_17, ciamei_22, ravensbergen_18, ye_2020, hansen_13, hara_11, pires_14}.
The mass imbalance separates the timescale of atomic collisions from that of thermalization, because the latter requires many more collisions due to the small energy transfer per collision.
In order to measure both timescales, we compress the Li atoms by ramping up the tune-out beam power from an adiabatically ramped initial value of \qty{10}{\milli\watt} to a fixed final one of \qty{30}{\milli\watt}.
This induces an anisotropic change in the trapping potential of Li.
The ramp is adiabatic on the single-particle timescale set by the trap frequencies, but constitutes a quench with respect to the collisional timescales.
The timescale of atomic collisions is observable as cross-dimensional relaxation (isotropization) of the cloud.
The duration of the quench is chosen such that a small but measurable amount of energy is inserted into the Li gas~\cite{supps}.
In~\autoref{fig:fig3}(b) and (d) we show an exemplary evolution of the reduced temperature and the anisotropy $\sigma_\text{rel} = (\sigma_z-\sigma_x)/(\sigma_z+\sigma_x)$ ($\sigma_{x,z}$ are the widths of the Li cloud measured in ToF) after the quench, from which we extract the respective $1/e$ timescales $\tau_{\text{th}}$ and $\tau_{\text{iso}}$ using exponential fits.
The bath temperature $T_\text{Er}$ remains constant, and in the absence of Er the quench-induced anisotropy persists on the experimental timescale due to the non-interacting nature of the fermions.
The latter constitutes the thermal-contact test introduced above.

We repeat this measurement across a range of Er bath temperatures from \qty{1.2}{\micro\kelvin} down to \qty{0.17}{\micro\kelvin}.
This allows us to extract both timescales while maintaining a relatively constant Fermi temperature $T_F$, which varies only slightly from \qty{6.7}{\micro\kelvin} to \qty{4.7}{\micro\kelvin} across the dataset~\cite{supps}.
In~\autoref{fig:fig3}(c) we analyze the dependence of the thermalization time $\tau_{\text{th}}$ quantitatively.
Importantly, $\tau_{\text{th}}\approx\qty{100}{\milli\second}$ remains constant over the explored range $0.06 \leq T/T_F \leq 0.18$.
This does not mean that Pauli blocking is absent: the average per-particle rate of ErLi collisions is suppressed $\propto T/T_F$, since only fermions within the thermally active shell around the Fermi energy $E_F$ find open final states.
This suppression is compensated exactly by the fermionic heat capacity, which is carried by the same shell and therefore scales as $T/T_F$ as well~\cite{presilla_03}.
In other words: Pauli blocking closes collision channels, but in the very same proportion it reduces the number of fermions whose excess energy has to be removed, leaving $\tau_{\text{th}}$ degeneracy independent.
Using $\tau_\text{th}^{-1} = \bar n\,\sigma_\text{ErLi}\,v_F\,\xi/3$ with the overlap density $\bar n$, the Fermi velocity $v_F$, the ErLi scattering cross section $\sigma_\text{ErLi} = 4\pi a_\text{ErLi}^2$, and the fraction of energy transferred per collision $\xi = 4 m_\text{Li} m_\text{Er}/(m_\text{Li} + m_\text{Er})^2 \approx 0.13$, we extract the magnitude of the background $s$-wave scattering length $|a_{\text{ErLi}}| = \qty{49 \pm 13}{\bohr}$, where $a_0$ is the Bohr radius~\cite{delannoy_2001, mudrich_02, supps}.
For our experimental parameters, the Er and Li clouds are nearly size-matched.
Consequently, the density overlap factor $\bar{n}$ remains near its maximum for the data shown in \autoref{fig:fig3}~\cite{supps}.
The thermalization time, corrected for the change in density overlap as $\tau_{\text{th}} \frac{\bar{n}}{\bar{n}_0}$, is shown by the light green circles in \autoref{fig:fig3}(c). The similarly corrected isotropization time $\tau_{\text{iso}} \frac{\bar{n}}{\bar{n}_0}$ is shown by the gray circles in \autoref{fig:fig3}(e), where $\bar{n}_0$ denotes the density overlap at the highest temperature.
Within our uncertainties, the two Li spin states are indistinguishable in their interaction with Er~\cite{supps}.
The repulsive character of this interaction, and thus the positive sign, $a_{\text{ErLi}} > 0$, is inferred from the directly visible density reduction appearing during ToF in the much faster expanding Li cloud (see inset of~\autoref{fig:fig3}(b)).
This scattering length of the $^6$Li-$^{166}$Er mixture is approximately half of that recently published for the $^7$Li-$^{166}$Er mixture~\cite{Kalia_26}.

Besides the absence of a slowdown of the thermalization with higher degeneracy, we also note that the thermalization is fast despite the large mass imbalance.
In comparison to a mass-balanced situation, the mass imbalance results in a reduced energy transfer per collision, measured by the fraction $\xi \approx 0.13$ introduced above.
This is often cited as a challenge for thermalization in mass-imbalanced ultracold mixtures~\cite{mosk_2001, mudrich_02, marzok_07}.
We want to add two remarks on this issue: (i) Compared to intraspecies collisions of the heavy component (required for thermalization in any evaporation of this species), the velocities involved in the heavy-light collisions are governed by the light partner and are much larger.
This increases the scattering rate, transforming the linear-in-$\xi$ penalty on the thermalization time into a $\sqrt{\xi}$ dependence~\cite{mosk_2001}.
Moreover, during sympathetic cooling with the heavy species as the coolant, the overall timescale is set by the coolant's evaporation dynamics and is unaffected by the interspecies mass penalty~\cite{luiten_1996, hung_2008, supps}.
(ii) Deep in the fermionic degenerate regime, the small energy and magnitude of momentum transfer per collision are, in fact, beneficial.
This ensures that Pauli blocking effects in the collisions with a very cold, but classical, bath of the heavy species remain small, because the majority of scattering events happen inside the thermally active shell~\cite{supps}.
Also for a degenerate bosonic bath, the large mass helps to keep thermalization efficient due to the supersonic nature of the collisions between the condensate and the light fermions~\cite{timmermans_1998, supps}.
Here the high mass imbalance keeps the Fermi velocity ($v_F\propto m_\text{Li}^{-3/4}$) well above the speed of sound ($v_c\propto m_\text{Er}^{-1}$) of the BEC.

The analysis of the dependence of the isotropization time on $T/T_F$ reveals a different behavior than that of the thermalization time (see~\autoref{fig:fig3}(e)).
We observe a clear slowing down, in agreement with a $(T/T_F)^{-1}$ scaling, and the match becomes even better when taking the changing density overlap into account.
The difference to the thermalization is the absence of the compensation mechanism: while Pauli blocking suppresses the collision rate $\propto T/T_F$ as before, the anisotropic quench deforms the entire Fermi sea, such that the fraction of atoms to be redistributed does not decrease with $T/T_F$.
The blocking therefore remains uncompensated and appears as the observed scaling $\tau_\text{iso}^{-1} \propto T/T_F$.
The same reasoning would apply to thermalization after a large-amplitude quench.

Isotropization, however, is particularly sensitive because the anisotropic change in trap frequencies directly affects the Fermi sea.
We illustrate both processes by the sketch in~\autoref{fig:fig3}(a).

We establish isothermal state preparation, independent of the external potential, as a new method for ultracold fermions, breaking with the central experimental paradigm of adiabatic preparation.
Our results shed light on the thermalization dynamics of deeply degenerate light fermions immersed in a bath of heavy particles:
thermalization remains effective even in the deeply degenerate regime, and the mass imbalance, often quoted as being detrimental for sympathetic cooling, is even beneficial for reaching record-low temperatures.
The isothermal preparation generalizes directly to more complex potentials or fermionic settings, such as reduced-dimensional systems, optical lattices, and strongly spin-imbalanced systems, where reaching deep degeneracy is notoriously difficult.
Whereas established entropy-removal schemes redistribute entropy within a specific lattice geometry into metallic wings or a band-insulator core~\cite{ho_09, bernier2009}, as demonstrated in Fermi-Hubbard experiments~\cite{xu_25}, or by entropy exchange in a species-selective lattice~\cite{catani_09}, or rely on dissipative superfluid immersion~\cite{griessner2006}, our approach transfers entropy to the independently controlled and potentially large bosonic reservoir.
Three independent experimental upgrades offer clear pathways to even deeper quantum degeneracies:
First, the thermalization time scale can be decreased by tuning near an interspecies Feshbach resonance~\cite{schafer2022-cg}.
Second, our scheme can be extended to the even narrower \qty{1299}{\nano\meter} transition of Er, which further reduces off-resonant scattering by more than an order of magnitude.
Third, the absolute temperature of the Er bath itself can be reduced by an adiabatic decompression into a shallower trapping potential.


\bigskip
\textit{Acknowledgments} -- We acknowledge stimulating discussions with Rudi Grimm. We acknowledge funding from the Horizon Europe program HORIZON-CL4-2022-QUANTUM-02-SGA via the project 101113690 (PASQuanS2.1), the Federal Ministry of Education and Research Germany (BMBF) via the project FermiQP (13N15889), the Deutsche Forschungsgemeinschaft within the research unit FOR5522 (Grant No. 499180199) and the Alfried Krupp von Bohlen and Halbach foundation.

\bigskip
\textit{Author Contributions} -- All authors contributed extensively to the planning, data acquisition, or analysis of the results presented here.

\bigskip
\textit{Data availability} --
The experimental and theoretical data and evaluation scripts that support the findings of this study will be available on Zenodo.

\bigskip
\textbf{Competing interests:}
There are no competing interests to declare.

\bibliography{bibliography}

\FloatBarrier

\newpage

\section{Supplemental Material}

\setcounter{figure}{0}
\renewcommand{\thefigure}{S\arabic{figure}}
\setcounter{equation}{0}
\renewcommand{\theequation}{S\arabic{equation}}

\subsection*{Detailed description of the mixture preparation}

The initial stages of the experiment, that is, the loading of the MOTs, proceed as described in~\cite{martino_25}.
Nevertheless, to make this supplement self-contained, we repeat the description of these steps here.
Our experiment features two individual, spatially separated MOT chambers and Zeeman slowers for Er and Li, both in line of sight with the glass cell.
The atomic beam for Er is derived from a commercial effusion cell oven and slowed down by a spin-flip Zeeman slower, utilizing Er's broad transition at \qty{401}{\nano\meter} ($\Gamma=2\pi\times\qty{29.7}{\mega\hertz}$).
After the slowdown process, Er is loaded into a narrow-line MOT at \qty{583}{\nano\meter} ($\Gamma=2\pi\times\qty{190}{\kilo\hertz}$)~\cite{ban_05, frisch_12}.
The low capture velocity required for the narrow-line MOT results in a large transverse spread of Er at the end of the Zeeman slower, leading to a reduced atom flux reaching the MOT.
Therefore, an intermediate slowdown of the atoms is performed shortly before the MOT region by the use of ``angled-slowing beams''~\cite{lunden_20}.
We reach a MOT loading rate of \qty{2e7}{\per\second}.

Simultaneously, the Li MOT is loaded, using the Li D$_2$-line at \qty{671}{\nano\meter} ($\Gamma=2\pi\times\qty{5.9}{\mega\hertz}$)~\cite{scherf_96}.
To increase the atom flux reaching the MOT, a combined transverse cooling and optical pumping stage is installed between the Li oven and the Zeeman slower.
Li is pumped into the hyperfine $\ket{F,m_F}=\ket{3/2,-3/2}$ state and simultaneously experiences a two-dimensional molasses on the D$_1$-line, resulting in a factor of \num{10} increased loading rate.
After a MOT loading time of \qty{5}{\second} and a consecutive cMOT stage, a Li cloud of \num{1e8} atoms is created at a temperature of about \qty{300}{\micro\kelvin} and in an equal spin mixture of $\ket{F,m_F}=\ket{1/2,\pm1/2}$.
Afterwards, Li is transferred to a crossed optical dipole trap, created by two horizontal beams with a waist of \qty{65}{\micro\meter}, overlapped with an angle of \ang{10}.
The light is derived from a multimode \qty{1070}{\nano\meter} laser, delivering a power of \qty{130}{\watt}.
Subsequently, a \qty{3}{\second} long evaporation at \qty{320}{\gauss} reduces the Li temperature and allows atoms to be loaded into the transport lattice, which is derived from two single-mode lasers operating at \qty{1064}{\nano\meter}.
Within \qty{55}{\milli\second} Li is then transported by \qty{0.5}{\meter} into the Er chamber.

Meanwhile, the Er MOT has been loading for \qty{8.5}{\second} and is now compressed, reaching a temperature of approximately \qty{8}{\micro\kelvin}, cold enough to be transferred directly into the transport lattice on top of Li.
Both species are then transported together by \qty{0.5}{\meter} into the glass cell within \qty{131}{\milli\second}.
To compensate gravity for Er, a magnetic field gradient of \qty{4.2}{\gauss\per\centi\meter} at a field of \qty{22}{\gauss} is created along the transport path.

Around the glass cell, three \qty{1064}{\nano\meter} beams derived from a single-mode laser and the tune-out beam are used for loading and evaporation.
For simplicity, only beam one (B1) and beam two (B2) are drawn in \autoref{fig:fig1}, as these two are used for evaporation.
For loading both species from the transport beam, we use B1 and beam three (B3) which is overlapped with B2.
B1 is overlapped with the transport beams and has a waist of $(w_h, w_v) = (29, 16)\,\unit{\micro\meter}$ and a power of \qty{2}{\watt}.
B3 is almost perpendicular to B1, has a waist of $(w_h, w_v) = (360, 43)\,\unit{\micro\meter}$ and an initial power of \qty{20}{\watt}.
After \qty{4}{\second}, both species are thermalized in B1 and B3.
At this point, we have \num{3e5} Er atoms and up to \num{2e4} Li atoms.
We observe a significant Er loss within the \qty{4}{\second} of thermalization, as Li is initially populating high-energy levels of the ODT provided by B1 and B3.
Hence, in the measurements presented here, we reduced the Li atom number to about \num{1e4} by decreasing the power of the Li MOT beams.
\begin{figure}
	\centering
	\includegraphics{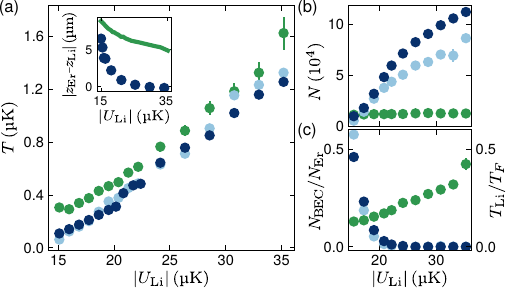}
	\caption{Sympathetic cooling of Li (green) via the evaporation of Er in a dual-species \qty{1064}{\nano\meter} optical dipole trap.
		The temperature evolution of the Er bath is shown both in the presence (light blue) and absence (dark blue) of Li.
		The trap depth for Li $U_\text{Li}$ stays finite when the trap has already opened for Er due to gravity at the end of the evaporation ramp.
		(a) Independently measured absolute Li and Er temperatures $T_\text{Li}$ and $T_\text{Er}$ match at the start of the evaporation ramp and split in the end due to Li imaging limitations discussed in this supplement.
		The inset shows the displacement of Er $|z_\text{Er}-z_\text{Li}|$ over the evaporation due to gravity, compared to the vertical Fermi radius of Li (green line).
		(b) Sympathetic cooling of Li is demonstrated by maintaining a constant atom number $N_\text{Li}\approx 10^4 $ during the evaporation, while Er atom numbers $N_\text{Er}$ are decreasing.
		The additional heat load on Er due to Li reduces $N_\text{Er}$.
		(c) Creation of a double-degenerate system is manifested in a non-vanishing condensate fraction $N_\text{BEC}/N_\text{Er}$ and a deeply degenerate Li temperature $T/T_F$, measured by fitting a finite temperature Fermi distribution (see~\autoref{equ:2D_Fermi_profile}) to Li.
	}
	\label{fig:figsi1}
\end{figure}
After the \qty{4}{\second} of thermalization, B1 is ramped to \qty{1}{\watt}, B3 is ramped down fully and B2 is ramped up to \qty{17}{\watt} within \qty{1}{\second}.
B2 has a waist of $(w_h, w_v) = (32, 225)\,\unit{\micro\meter}$ and significantly increases the horizontal trap frequency for Er and Li.
The subsequent evaporation is split into two linear power ramps.
Within the first evaporation ramp B1 is ramped down to \qty{0.3}{\watt} and B2 is ramped down to \qty{2}{\watt} within \qty{0.35}{\second}, reducing the trap depth for Li from \qty{178}{\micro\kelvin} to \qty{36.4}{\micro\kelvin}.
During the second evaporation ramp, B1 is ramped down to \qty{0.07}{\watt} within \qty{4.3}{\second}, further reducing the trap depth for Li to \qty{15}{\micro\kelvin}.
While the Er atom number is decreasing during the evaporation ramp, the Li atom number is constant (see \autoref{fig:figsi1}b).
We measured the evaporation trajectory of Er in the presence and absence of Li (see \autoref{fig:figsi1}a).
Especially at the start of the evaporation there is a significant difference due to the additional heat load when loading Li from the transport lattice into the glass cell traps.
For deeper evaporation times, a BEC of Er is observed, while Li is also degenerate with a temperature of $T/T_F=0.14$ (see \autoref{fig:figsi1}c).
This number, measured directly on the fermions, is strongly influenced by the imaging limits discussed below.
Using the bosons as a thermometer results in $T/T_F=0.06$.
The detection-induced larger relative difference towards lower temperatures is apparent.
At the final evaporation point, the gravitational sag of Er is almost equal to the Fermi radius of Li (see inset of~\autoref{fig:figsi1}(a)).

\subsection*{Description of the dual-species imaging}
We use standard absorption imaging to determine the atomic column density distributions of both species.
The drastically different expansion velocities of Li and Er allow us to image the two species sequentially after their release from the optical trap.
The imaging sequence begins with the fast-expanding Li cloud.
After release, Li is in a balanced mixture of the $|1\rangle$ and $|2\rangle$ hyperfine spin states, with the quantization axis defined by a magnetic field oriented along the $z$-direction.
During the \qty{1}{\milli\second} time-of-flight, the magnetic field is rapidly adjusted to $B = \qty{3.6}{\gauss}$, oriented parallel to the imaging axis (see \autoref{fig:fig1}).
Then we apply a \qty{10}{\micro\second} optical pumping pulse resonant on the D$_1$-line, driving the $^2S_{1/2} |F=1/2\rangle \to {}^2P_{1/2} |F'=3/2\rangle$ transition.
The optical pumping light is directed counter-propagating to the imaging beam.
Keeping the pumping light, we apply a \qty{20}{\micro\second} imaging pulse resonant with the D$_2$-line, targeting the $^2S_{1/2} |F=1/2\rangle \to {}^2P_{3/2} |F'=3/2\rangle$ cycling transition.
The Li density is visibly reduced locally in the region of the Er cloud (see inset of \autoref{fig:fig3}), which signals a repulsive interaction between the two species.
Furthermore, the measured Li atom number $N_{\text{Li}}$ may be subject to a systematic underestimation of up to $\qty{50}{\percent}$ due to imperfect optical pumping and non-closure of the imaging transition.

Er imaging is performed following the Li detection sequence.
After an additional period of \qty{5}{\milli\second} time-of-flight expansion, a \qty{200}{\micro\second} imaging pulse addressing the \qty{401}{\nano\meter} line is applied along the same optical axis used for the Li imaging beam (see \autoref{fig:fig1}(a)).
By the time the Er images are taken, the Li density has become negligibly small due to its rapid expansion and does not influence the Er density profile.
The Li and Er imaging beams are separated by a dichroic mirror and detected on separate, dedicated CCDs.

\subsection*{Detection limit of direct $T/T_F$ fitting}
\begin{figure}
	\centering
	\includegraphics{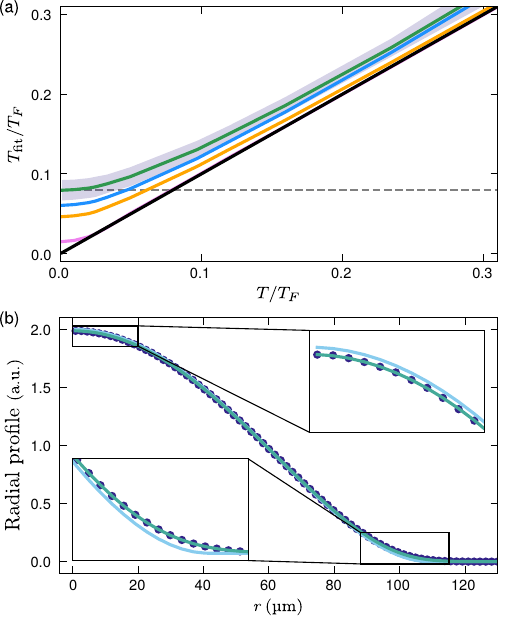}
	\caption{(a) Simulated impact of individual imaging noise sources on the accuracy of temperature extraction.
		Measured temperature $T_\text{fit}/T_F$  by fitting a fermionic density distribution (\autoref{equ:2D_Fermi_profile}) at $T/T_F$ (black) with limited resolution (pink), center fluctuations (orange), atom number fluctuations (blue) and the convoluted effect of all three (green).
		The gray region represents the standard deviation of additional pixel noise for 50 shot averaging.
		The lowest measurable reduced temperature of $T/T_F=0.08$ is indicated by the dashed line.
		(b) Azimuthal average of the simulated noise free density profile at $T/T_F=0$ (light blue) in comparison to a simulated noise affected profile (blue dots), and the least squares fit (green) giving $T/T_F=0.08$.
	}
	\label{fig:figsi2}
\end{figure}
For degenerate fermionic gases below $T/T_F \approx 0.5$, the temperature affects only the shape and not the size, which is dominated by the Pauli pressure.
To determine the temperature, we apply a least-squares fit of the 2D density profile of a non-interacting Fermi gas \cite{ketterle_08} in \autoref{equ:2D_Fermi_profile} to the absorption images
\begin{equation}
	\label{equ:2D_Fermi_profile}
	n(x,y) = n_0 \frac{\mathrm{Li}_2 \left( - \exp\left[q - \left(\frac{x^2}{R_x^2} + \frac{y^2}{R_y^2} \right) f(e^q)\right] \right)}{\mathrm{Li}_2(-e^q)}
\end{equation}
where $R_{x,y}$ are the Fermi radii in the given direction, $\mathrm{Li}_n$ is the polylogarithm of order $n$, $f(x)=\frac{1+x}{x}\ln(1+x)$ and $q=\mu\beta$ is the logarithm of the fugacity.
The parameter $q$ determines the shape and thus the temperature of the gas through the relation
\begin{equation}
	\frac{T}{T_F} = \left[-6\, \mathrm{Li}_3(-e^q)\right]^{-1/3}.
\end{equation}
As the parameter $q$ is only given by the shape of the cloud, the quality of the absorption images and the stability of the system have a large effect on the detectable temperature using this method.
Multiple effects were considered in the evaluation of the limitation to the detectable temperature due to cloud distortion.
The effects that are straightforward to quantify in numerical simulations include the finite resolution of the imaging system, pixel readout noise, atom number fluctuations and the consequential change in the Fermi radius, and the shot-to-shot position fluctuation of the cloud.
The latter two effects are relevant only for absorption images averaged over multiple repetitions, which is necessary for our system.
In addition to these effects, there are further distortions on the absorption image caused by interference fringes, mean-field repulsive forces during the time of flight and forces from the repumper light before and during the imaging, which are not quantitatively investigated here.

For the investigation of the detection limit, the strength of each effect is determined from absorption images of the degenerate Li cloud.
50 images are simulated for each temperature using \autoref{equ:2D_Fermi_profile} before they are averaged and the fit is applied.
To each image normally distributed random noise on the atom number (Fermi radius), center position and individual pixel values has been added.
The resolution of the simulations reflects our effective pixel size of \qty{2.15}{\micro\meter}.
The amplitude of the simulated Fermi profile was chosen to reflect the peak column density of the Li gas in our absorption images with a peak optical density of two.
The expanded Fermi radius after \qty{1}{\milli\second} ToF of our round cloud was determined to be $R_F = \qty{109}{\micro\meter}$ and is scaled in the simulations by $(N/N_0)^{1/6}$ to simulate the effect of atom number fluctuations in $N$ with $\sigma_N=\qty{30}{\percent}$.

The pixel noise was simulated with normally distributed noise around 0 with $\sigma_n=1.28$ added on top of each simulation.
For cloud center fluctuations, normally distributed random offsets in $x$ and $y$ were used with $\sigma_\text{center} = 2.2$~pixels or $\sigma_\text{center}\approx\qty{4.7}{\micro\meter}$.

The effect of the individual noise sources on the detectable temperature is shown in \autoref{fig:figsi2}(a).
The noise added to the pixels caused the fitted temperature to fluctuate for each repetition of the 50 simulated shots as indicated by the shaded area defined by one standard deviation of the fluctuating temperature fit.
\autoref{fig:figsi2}(b) shows the distortion of the simulated density profiles and demonstrates how small changes in the cloud's shape affect the determined temperature.

\subsection*{Vertical alignment and spatial overlap}
\begin{figure}
	\centering
	\includegraphics{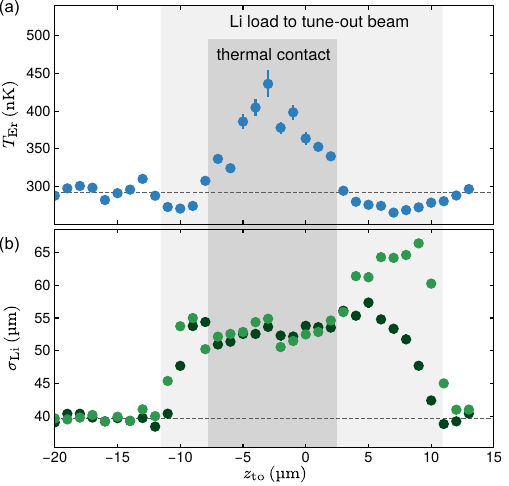}
	\caption{Vertical alignment of the tune-out beam.
		The coordinate $z_{\text{to}} = 0$ is defined as the vertical Li cloud position in the absence of the tune-out beam.
		The black dashed line shows the value without the tune-out beam.
		(a) Er temperature $T_{\text{Er}}$ as a function of the vertical position of the tune-out beam.
		The peak in $T_{\text{Er}}$ at $z_{\text{to}} \approx \qty{-3}{\micro\meter}$ indicates the optimal spatial overlap between the Er and Li clouds.
		(b) Horizontal (dark green) and vertical (light green) widths after ToF of the Li cloud for different tune-out beam positions.
	}
	\label{fig:figsi3}
\end{figure}
The tune-out beam is elliptical, with waists of $(w_h, w_v) = (22, 6)\,\unit{\micro\meter}$.
Its polarization is linear and orthogonal to the quantization axis defined by the magnetic field.
Due to the anisotropic polarizability of Er, this configuration maximizes the detuning from the nearby narrow resonance~\cite{ban_05, martino_25}.
Species-selective trapping enables us to precisely compensate for the differential gravitational sag arising from the extreme mass imbalance between Er and Li.
In \autoref{fig:figsi3}(a) and (b) we show the effect of the changing tune-out beam position on Er temperature $T_\text{Er}$ and width of the Li cloud $w_\text{Li}$.
We define the reference $z_{\text{to}} = 0$ as the vertical position of the Li cloud in the absence of the tune-out beam.
A maximum in $T_{\text{Er}}(z_{\text{to}})$ is observed when the tune-out beam is positioned \qty{3}{\micro\meter} below this initial reference.
At this optimal overlap point, the Li cloud width $w_{\text{Li}}$ is larger than the width measured without the tune-out beam (indicated by the black dashed line).
This expansion is a direct consequence of the increased Fermi energy $E_F$ within the tune-out potential.
Spatial overlap between the Er and Li clouds is maintained within the range $\qty{-8}{\micro\meter} < z_{\text{to}} < \qty{3}{\micro\meter}$.
For vertical positions $\qty{3}{\micro\meter} < z_{\text{to}} < \qty{9}{\micro\meter}$, the Li cloud is successfully loaded into the tune-out trap but remains spatially isolated from the Er reservoir.
This results in $T_{\text{Er}}$ values lower than without the tune-out beam.
Furthermore, the Li cloud exhibits significant anisotropy in this region, highlighting the lack of sufficient intraspecies rethermalization for Li without the presence of Er.
In the regions $\qty{-20}{\micro\meter} < z_{\text{to}} < \qty{-12}{\micro\meter}$ and $\qty{12}{\micro\meter} < z_{\text{to}} < \qty{15}{\micro\meter}$, neither the Er nor the Li clouds are affected by the tune-out beam.
At these positions, the beam does not overlap with the initial Li cloud, preventing any atoms from being loaded into the tune-out potential.
\autoref{fig:figsi3} illustrates the high level of control achieved over the relative position of the Li and Er clouds.
The differential gravitational sag depends on the depth of the Er potential $U_{\text{Er}}$, which changes throughout the evaporation sequence.
In the experimental setup presented here, we lack the ability to adjust the tune-out beam position dynamically during evaporation.
While the gravitational sag can be fully compensated for any specific trap depth, implementing dynamic adjustment of the tune-out beam could further optimize both the loading and cooling processes in the future.

\subsection*{ErLi mixture stability}
\begin{figure}
	\centering
	\includegraphics{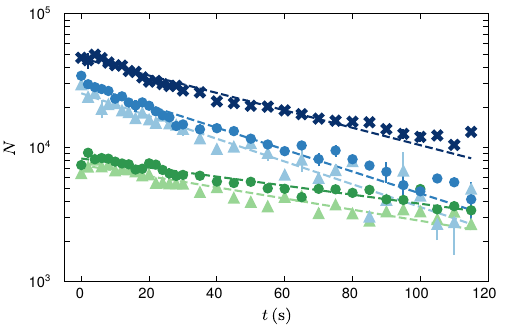}
	\caption{Lifetime measurements of the ErLi mixture for various spin-state combinations.
		The dotted lines represent exponential fits used to extract the characteristic $1/e$-lifetimes:
		Er alone: \qty{67 \pm 3}{\second} (dark blue crosses), Er \qty{53 \pm 2.3}{\second} (blue circles) and Li $\ket{1}$ \qty{129 \pm 7}{\second} (green circles), Er \qty{51 \pm 3}{\second} (light blue triangles) and Li $\ket{1}\ket{2}$ \qty{104 \pm 6}{\second} (light green triangles).
	}
	\label{fig:figsi4}
\end{figure}

We investigated the collisional stability of the different ErLi mixtures by recording the evolution of the atom numbers for varying hold times at the beginning of the evaporation ramp (see \autoref{fig:figsi4}).
We studied three distinct cases: Er alone, Er with Li spin-polarized in $\ket{1}$, and Er with a balanced Li spin mixture of states $\ket{1}$ and $\ket{2}$.
These measurements were performed at a temperature of $T \approx \qty{1.5}{\micro\kelvin}$.
A spin-polarized sample of Li in state $\ket{1}$ is prepared by applying a Landau-Zener sweep to transfer the population from $\ket{2}$ to $\ket{3}$.
Subsequently, a resonant light pulse is applied driving the cycling transition and consequently removing $\ket{3}$ population, leaving a pure $|1\rangle$ sample.
The initial Li atom number is adjusted to have identical total atom number for both the polarized and balanced configurations.
In the absence of Li, the Er $1/e$-lifetime is found to be \qty{67 \pm 3}{\second}.
The presence of about \num{e4} Li atoms leads to a similar modest reduction of the Er lifetime, yielding values of \qty{53 \pm 2.3}{\second} for the polarized mixture and \qty{51 \pm 3}{\second} for the balanced configuration.
Furthermore, we observe a reduction in the initial Er atom number when Li is present in the trap due to the discussed initial loading mechanisms.
The $1/e$-lifetimes for Li were determined to be \qty{129 \pm 7}{\second} for the spin-polarized configuration and \qty{104 \pm 6}{\second} for the balanced mixture.
The small differences in the Er and Li lifetimes between the two mixtures, combined with their long absolute values, demonstrate the high collisional stability of the system even at low field.
Losses from dipolar relaxation of Li $\ket{2}$ to $\ket{1}$ in the vicinity of Er prove to be negligible for this configuration.
Furthermore, neither the thermalization measurements presented in the main text nor sympathetic cooling measurements with a single-component Li gas in either state $\ket{1}$ or $\ket{2}$ allow us to distinguish the collisional properties of the two spin states with Er within our uncertainties.

Close inspection of the data in~\autoref{fig:figsi4} reveals that the data are not linear on the semilogarithmic plot.
This is expected, as the atom number loss is dominated by two- or three-body loss processes.
Extracting the exact rate coefficients goes beyond the scope of this study and we used the exponential fits reported above to estimate the timescales of the decay in our specific setting.
The main goal is to show that mixing Er and Li does not lead to extensive losses.

\subsection*{Interspecies thermalization}

\begin{figure*}
	\centering
	\includegraphics{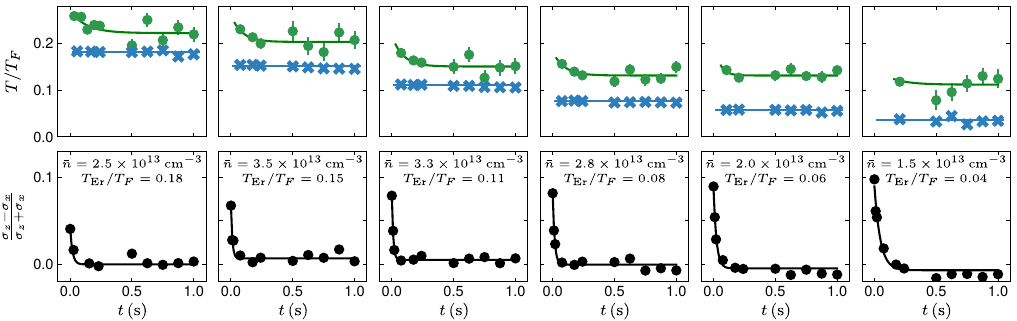}
	\caption{Temperature relaxation of Er (blue) and Li (green), and isotropization of Li (black), after a quench on Li for various initial temperatures.}
	\label{fig:figsi5}
\end{figure*}

The temperature relaxation of both species and the isotropization of Li, from which we extract the timescales $\tau_\text{th}$ and $\tau_\text{iso}$, are shown in \autoref{fig:figsi5} for the full range of explored temperatures.
Initially, Li has been selectively heated by a quench with the tune-out beam.
The total collision rate between Er and Li is given by: $\Gamma_\text{coll}^\text{tot}=\Gamma\cdot \pi^2T_\text{Li}/(3T_F)$.
The (classical) collision rate is given by $\Gamma = \bar n_{\text{ErLi}} \sigma \bar v$ and the factor $\pi^2T_\text{Li}/(3T_F)$ accounts for the reduction in available states due to Pauli blocking.
The number of collision partners per unit volume $\bar n_{\text{ErLi}}=\int n_\text{Er}(\textbf{r}) n_\text{Li}(\textbf{r})d^3\textbf{r}$ is obtained from the thermal density distribution $n_\text{Er}$ for Er and we approximate $n_\text{Li}$ with the zero temperature Fermi density distribution for Li.
The interspecies thermalization cross section $\sigma_{\text{ErLi}}=4\pi a_{\text{ErLi}}^2$ is determined by the $s$-wave scattering length $a_{\text{ErLi}}$.
The mean relative velocity $\bar v=v_F$ is well approximated by the Fermi velocity $v_F$ of Li, since the thermal velocity of Er is negligible against $v_F$.
The energy transfer per ErLi collision $\Delta E= \xi k_B\Delta T$ is proportional to the temperature difference $\Delta T=T_\text{Li}-T_\text{Er}$, with the factor $\xi = 4m_\text{Li}m_\text{Er}/(m_\text{Li}+m_\text{Er})^2\approx 0.13$ accounting for the reduction in energy transfer due to the unequal masses $m_i$~\cite{mudrich_02}.
The total heat transfer rate $\dot Q=\Gamma^\text{tot}_\text{coll}\Delta E$ changes the total internal energy $\dot {\mathcal{U}}_i$ of the two gases $ \dot Q= \dot {\mathcal{U}}_\text{Er}=- \dot {\mathcal{U}}_\text{Li}$ in our isolated system.
Using the Sommerfeld expansion for $T\ll T_F$, the total internal energy of Li is given by ${\mathcal{U}}_\text{Li}(T_\text{Li})=\frac{3}{4}N_\text{Li} k_B T_F + \frac{\pi^2}{2}N_\text{Li}k_BT_F\left( \frac{T_\text{Li}}{T_F}\right)^2 + \mathcal{O}(T_\text{Li}^4)$.
The energetic dynamics for Li: $\dot {\mathcal{U}}_\text{Li}=C_\text{V,Li}(T_\text{Li})\dot T_\text{Li}$ is determined by the temperature-dependent heat capacity $C_\text{V,Li}=\partial \mathcal{U}_\text{Li}/\partial T_\text{Li}= \pi^2 k_B N_\text{Li} \frac{T_\text{Li}}{T_F}$.
We obtain the thermally active fraction of Li by comparing the heat capacity of the degenerate system to that of a classical gas ($3k_BN_\text{Li}$): $C_\text{V,Li}/3k_BN_\text{Li}=\pi^2/3\cdot T/T_F$.
The total internal energy of Er is given by $ {\mathcal{U}}_\text{Er}=C_\text{V,Er} T_\text{Er}$ with the classical temperature independent heat capacity $C_\text{V,Er} =3k_BN_\text{Er}$ and with this: $\dot {\mathcal{U}}_\text{Er}=C_\text{V,Er} \dot T_\text{Er}$. The temperature dynamics
\begin{align}
	\frac{d}{dt}\Delta T & =\!\frac{\dot {\mathcal{U}}_\text{Li}}{C_\text{V,Li}}\!-\!\frac{\dot {\mathcal{U}}_\text{Er}}{C_\text{V,Er}}\nonumber                                                                              \\
	                     & =\!-\left[\! \frac{\bar n_{\text{ErLi}}  \sigma_{\text{ErLi}} v_F \xi}{3 N_\text{Li}}\! \left(\! 1\!+\!\pi^2\frac{N_\text{Li} T_\text{Li}}{N_\text{Er}T_F}\! \right)\! \right]\! \Delta T\nonumber \\ &\approx -\frac{\bar n_{\text{ErLi}} \sigma_{\text{ErLi}} v_F\xi}{3 N_\text{Li}}\Delta T
	\label{equ:themp_dynamics}
\end{align}
follows an exponential with the relaxation rate
\begin{equation}
	1/\tau_\text{th}=\frac{\bar n \sigma_{\text{ErLi}} v_F\xi}{3},
	\label{equ:s-wave1}
\end{equation}
where $\bar n= \bar n_{\text{ErLi}}/N_{\text{Li}}$ is the overlap density. Since $N_\text{Li}/N_\text{Er}\ll1$ and $T/T_{F}<1$, the temperature-dependent contribution to the relaxation is dropped.
The obtained relaxation time is not dependent on $T/T_F$, which keeps sympathetic cooling with Er efficient, since the temperature-dependent Pauli blocking of available states and the temperature dependence in the heat capacity for fermions cancel exactly.
This cancellation does not arise by construction, but from a consistent definition of the thermally active fraction of the fermions.

In the scenario of Li as a thermal gas, the relaxation rate similarly reads $1/ \tilde \tau_\text{th}=\frac{\bar n \sigma_{\text{ErLi}} v_\text{th}\xi}{3}$, with the mean relative thermal velocity $v_\text{th}=\sqrt{8k_B(T_\text{Er}/m_\text{Er} + T_\text{Li}/m_\text{Li})/\pi}\approx\sqrt{8k_BT_\text{Li}/m_\text{Li}/\pi}$ and for $\bar n$ a thermal distribution for Li is considered.
Having much less Li than Er and performing the thermalization for a degenerate Fermi gas helps to maintain constant density overlap $\bar n$ and mean relative velocity during the experiment.

We extract the interspecies scattering length $a_{\text{ErLi}}$ from the measured thermalization time $\tau_{\text{th}}$ via an exponential fit to the Li temperature relaxation data shown in \autoref{fig:figsi5}.
This analysis explicitly accounts for the changing interspecies density overlap $\bar{n}$ and the Fermi velocity $v_{\text{F}}$.

We further note that the reduced energy transfer per collision, quantified by $\xi$, becomes irrelevant during sympathetic cooling whenever the heavy component is the coolant.
There, the required timescale is dominated by the exponential factor in the evaporation timescale $\tau_\text{evap} \approx \tau_\text{th,Er}\exp(\eta)$~\cite{luiten_1996, hung_2008}, with typical ratios of the trap depth to the temperature of $\eta \approx 8$.
The intraspecies thermalization time of the coolant (here $\tau_\text{th,Er}$) sets this scale, and the small mass penalty increasing $\tau_\text{th,ErLi}$ has no effect.

\subsection*{Temperature offset in steady state}
In our experiment, the dominant heating source for Li is off-resonant photon scattering from the tune-out trap light at \qty{841}{\nano\meter}, which is primarily driven by the \qty{671}{\nano\meter} Li transition.
Since the scattering rate is proportional to the local light intensity, we normalize it to the Li trap depth $U_{\text{Li}}$ generated by the tune-out beam.
Given a large detuning-to-linewidth ratio of $\Delta_{671}/\Gamma_{671} \approx \num{1.5e7}$, we obtain a
scattering rate of $\Gamma_\text{sc}/U_{\text{Li}} = \qty{4.9}{\milli\hertz\per\micro\kelvin}$~\cite{grimm_99} and a corresponding heating rate of $\gamma_{\text{heat}}=\frac{2}{3}\Gamma_\text{sc}E_{\text{rec}}/k_{\text{B}}\approx \qty{7.3}{\nano\kelvin\per\second} \cdot U_{\text{Li}}/\unit{\micro\kelvin}$, where $E_{\text{rec}}=\hbar^2k^2/2m$ is the recoil energy.
Due to its significantly lower recoil energy, the corresponding heating rate for Er is negligible.
This continuous heating of Li modifies the temperature dynamics described in \autoref{equ:themp_dynamics}:
\begin{equation}
	\frac{d}{dt}\Delta T = - \Delta T/\tau_\text{th} +\gamma_{\text{heat}}.
	\label{equ:themp_heating}
\end{equation}
Under steady-state conditions ($\frac{d}{dt}\Delta T = 0$), this constant heating yields a residual temperature offset between the two species of $\Delta T = \gamma_{\text{heat}} \tau_{\text{th}}$.
Consequently, this additional heating term may lead to a systematic underestimation of the true fermion temperature when relying solely on bosonic thermometry.
At the lowest achieved Er temperature of $T_{\text{Er}} = 0.024~T_F$, with a trap depth of $U_{\text{Li}}=\qty{49}{\micro\kelvin}$ and the mean thermalization time of $\tau_\text{th}\approx\qty{100}{\milli\second}$, this effect generates a temperature offset of $\Delta T \approx \qty{36}{\nano\kelvin} \approx 0.007~T_F$.
This yields the upper bound of the deepest achieved Li degeneracy of $T_{\text{Li}}/T_F = (T_{\text{Er}} + \Delta T)/T_F = \num{0.024}^{+0.007}$ quoted in the main text.

\subsection*{Kinematic bypass of superfluid suppression of the scattering rate}
The following arguments are discussed in detail in~\cite{timmermans_1998}.
Here we summarize them for the sake of completeness of the heavy-light thermalization discussion.
In a BEC, the linear phonon dispersion at low momenta significantly modifies the excitation spectrum, leading to a substantial suppression of the impurity scattering rate.
The total impurity scattering rate $\Gamma$ is determined by the dynamic structure factor $S(\mathbf{q}, \omega)$ via Fermi's Golden Rule:
\[\Gamma = \frac{V}{(2\pi)^3} \int d^3q \left[ \frac{2\pi}{\hbar} |v_{\mathbf{q}}|^2 S(\mathbf{q}, \omega_{\mathbf{q}}) \right],\]
where $v_{\mathbf{q}}$ is the Fourier transform of the external perturbation.
Because $S(\mathbf{q}, \omega)$ is strongly suppressed in the long-wavelength, collective phonon regime, one might expect a significant slowdown of the thermalization rate for an impurity interacting with the condensate; in particular, when moving at velocities below or near the Bogoliubov sound velocity $c$.
However, for the specific case of an ErLi mixture, this suppression effect is negligible.
This bypass is driven by two key factors: the extreme difference in atomic masses ($m_{\text{Li}} \ll m_{\text{Er}}$) and the differential trapping potentials applied to the two species. These physical conditions dictate that the velocity of the Li atoms is intrinsically much higher than the speed of sound in the Er BEC ($v_{\text{Li}} \gg c$).
Consequently, the available momentum transfer $q$ is pushed far beyond the collective excitation limit, forcing the interactions deep into the free-particle regime.

\subsection*{Light(fermion)-heavy collision suppression}
The large mass imbalance between Er and Li reduces the energy transfer in a single collision event, as discussed in the main text body.
The maximum relative energy transfer is given by $\Delta E/E=\xi=\frac{4 m_\text{Li}m_\text{Er}}{(m_\text{Li}+m_\text{Er})^2}$.
Collisions do not suffer from Pauli blocking if the energy transfer stays within the thermally active shell.
At low temperatures the energy per particle in the Fermi gas is given by $\mathcal{U}_\text{Li}/N_\text{Li}=\epsilon_0(T=0) + \frac{\pi^2}{2}E_F(T/T_F)^2$ in the Sommerfeld approximation.
The temperature dependent term serves as a measure for the width of the thermally active shell $\Delta E_\text{th}/E_F= 2\cdot \frac{\pi^2}{2}(T/T_F)^2$.
The factor "$2$" accounts for atoms being cooled from above the Fermi energy to below the Fermi energy.
With the condition $\Delta E /E<\Delta E_\text{th}/E_F$, an expression for the temperature at which the energy transfer becomes comparable to the width of the thermal shell is obtained:

$$\frac{T}{T_F}>\frac{\sqrt{\xi}}{\pi}\approx 12\%.$$

The $s$-wave collision process spans all energies up to the scale set by $\xi E$.
Thus, only significantly below this temperature, the effect of Pauli blocking starts to become relevant.

\subsection*{Linear fits to the compression data}

While \autoref{fig:fig2}(b) displays the Er temperature as a function of $T_F$, we extract the Er heating rate by applying a linear fit to the same dataset evaluated against the Li trap depth, $T_{\text{Er}}(U_{\text{Li}})$.
From highest to lowest initial temperature, fits to the mixture data yield $T_{\text{Er}} = 1.18(6) + 0.0017(2)\, U_{\text{Li}}$ (light blue circles), $T_{\text{Er}} = 0.536(3) + 0.0010(1)\, U_{\text{Li}}$ (blue circles), and a constant $T_{\text{Er}} = 0.1159(7)$ (dark blue circles). The control data without Li (crosses) exhibit no dependence on $U_{\text{Li}}$, yielding constant baselines of $T_{\text{Er}} = 0.9697(37)$ (light blue), $0.4299(24)$ (blue), and $0.1248(3)$ (dark blue). All temperatures and trap depths are given in \unit{\micro\kelvin}.
The exact magnitude of the Er temperature increase in the presence of Li is set by the energy flow from the compressed fermions and depends on the Er atom number and the total Li heat capacity.
Additionally, we extract the Er loss rate by applying a linear fit to the atom number as a function of the Li trap depth, $N_{\text{Er}}(U_{\text{Li}})$.
For the Er atom numbers shown in \autoref{fig:fig2}(c), in units of $10^4$,
the mixture data (circles) yield $N_{\text{Er}} = 6.31(3)(1 - 0.00098(12)\, U_{\text{Li}})$ (light blue), $N_{\text{Er}} = 3.83(1)(1 - 0.0014(1)\, U_{\text{Li}})$ (blue), and $N_{\text{Er}} = 0.581(5)(1 - 0.0056(2)\, U_{\text{Li}})$ (dark blue).
The linear fits to the control data without Li yield $N_{\text{Er}} = 7.92(5)(1 - 0.00056(15)\, U_{\text{Li}})$ (light blue crosses), $N_{\text{Er}} = 5.15(3) (1- 0.00117(16)\, U_{\text{Li}})$ (blue crosses), and $N_{\text{Er}} = 1.018(4)(1 - 0.00181(9)\, U_{\text{Li}})$ (dark blue crosses).
The value reported above represents the effective loss rate, determined from the total atom loss over the combined 1.5-second sequence (a \qty{1}{\second} ramp-up followed by a \qty{0.5}{\second} hold).
Consequently, the time-averaged loss rate per second is two-thirds of the effective loss value.
In the compression data, $U_{\text{Li}}$ is varied from \qtyrange{0}{100}{\micro\kelvin}.

\subsection*{Heating-free atom loss of erbium}

The data shown in~\autoref{fig:fig2}(b, c) displays a Li-independent Er atom loss, which scales approximately linearly in the power of the tune-out beam and which does not induce measurable heating.
The observed loss rate is about $R_\text{loss,Er}/U_\text{Li} \approx \qty{2.5e-4}{\per \second \per \micro \kelvin}$, where we cannot distinguish between single and multi-atom processes based on the available data.
There are two processes that can explain this signal: (i) Raman scattering to different Zeeman states and subsequent dipolar collisions ejecting atoms from the trap and (ii) decay via long-lived dark states after photon scattering, such that the dark states leave the trap before the population reaches the ground state.

The total photon scattering rate of Er is $R_\text{scatt,Er}/U_\text{Li} \approx \qty{8e-3}{\per \second \per \micro \kelvin}$~\cite{martino_25}, with the observed loss amounting to \qtyrange{4}{15}{\percent} of this rate (where the lower bound corresponds to larger Er atom numbers).
This is reasonably explained by the two processes above, in particular, when taking into account that, after a spin flip, a dipolar collision converts about $\qty{100}{\micro \kelvin}$ of magnetic Zeeman energy into kinetic energy.
This is much larger than the trap depth, such that the two collision partners, and possibly even more atoms in secondary collisions, are removed from the trap.
In contrast to alkali atoms, Raman scattering does not switch off at large detunings for Er in its electronic ground state due to its tensor polarizability~\cite{lekien2013-cg,lepers2014-cg}.

\end{document}